\documentclass[12pt]{article}
\usepackage[pdftex]{graphicx}
\usepackage{xspace,colortbl}
\usepackage{amsmath}
\usepackage{slashed}
\usepackage[T1]{fontenc}
\usepackage[utf8]{inputenc}
\usepackage{authblk}
\date{}
\title{Decaying Vector Dark Matter as an  Explanation for the 3.5 keV Line from Galaxy Clusters}
\author[a]{Yasaman Farzan}
\author[a,b]{Amin Rezaei Akbarieh}
\affil[a]{ \small{School of physics, Institute for Research in
Fundamental Sciences (IPM)\\P.O.Box 19395-5531, Tehran, Iran\\}}
\affil[b]{Department of Physics, Sharif University of Technology,
P.O.Box 11155-9161, Tehran, Iran }

\begin{document}
\def\d{{\rm d}}
\def\Epos{E_{\rm pos}}
\def\ap{\approx}
\def\eff{{\rm eft}}
\def\L{{\cal L}}
\newcommand{\vev}[1]{\langle {#1}\rangle}
\newcommand{\CL}   {C.L.}
\newcommand{\dof}  {d.o.f.}
\newcommand{\eVq}  {\text{EA}^2}
\newcommand{\Sol}  {\textsc{sol}}
\newcommand{\SlKm} {\textsc{sol+kam}}
\newcommand{\Atm}  {\textsc{atm}}
\newcommand{\Chooz}{\textsc{chooz}}
\newcommand{\Dms}  {\Delta m^2_\Sol}
\newcommand{\Dma}  {\Delta m^2_\Atm}
\newcommand{\Dcq}  {\Delta\chi^2}
\newcommand{\nbb}{$\beta\beta_{0\nu}$ }
\newcommand {\be}{\begin{equation}}
\newcommand {\ee}{\end{equation}}
\newcommand {\ba}{\begin{eqnarray}}
\newcommand {\ea}{\end{eqnarray}}
\newcommand*{\Scale}[2][4]{\scalebox{#1}{$#2$}}%
\newcommand*{\Resize}[2]{\resizebox{#1}{!}{$#2$}}%
\renewcommand{\thefootnote}{\alph{footnote}}
\def\VEV#1{\left\langle #1\right\rangle}
\let\vev\VEV
\def\e6{E(6)}
\def\10{SO(10)}
\def\21{SA(2) $\otimes$ U(1) }
\def\321{$\mathrm{SU(3) \otimes SU(2) \otimes U(1)}$ }
\def\lr{SA(2)$_L \otimes$ SA(2)$_R \otimes$ U(1)}
\def\422{SA(4) $\otimes$ SA(2) $\otimes$ SA(2)}
\newcommand{\AHEP}{%
School of physics, Institute for Research in Fundamental Sciences
(IPM)\\P.O.Box 19395-5531, Tehran, Iran\\

  }
\newcommand{\Tehran}{%
School of physics, Institute for Research in Fundamental Sciences
(IPM)
\\
P.O.Box 19395-5531, Tehran, Iran}
\def\roughly#1{\mathrel{\raise.3ex\hbox{$#1$\kern-.75em
      \lower1ex\hbox{$\sim$}}}} \def\lsim{\roughly<}
\def\gsim{\roughly>}
\def\ltap{\raisebox{-.4ex}{\rlap{$\sim$}} \raisebox{.4ex}{$<$}}
\def\gtap{\raisebox{-.4ex}{\rlap{$\sim$}} \raisebox{.4ex}{$>$}}
\def\lsim{\raise0.3ex\hbox{$\;<$\kern-0.75em\raise-1.1ex\hbox{$\sim\;$}}}
\def\gsim{\raise0.3ex\hbox{$\;>$\kern-0.75em\raise-1.1ex\hbox{$\sim\;$}}}

\maketitle
\begin{abstract}
We present a Vector Dark Matter (VDM) model that explains the 3.5 keV line recently observed in the XMM-Newton observatory data from galaxy clusters.
 In this model, dark matter is composed of two vector bosons, $V$ and $V^\prime$, which couple to the photon through an effective
  generalized Chern-Simons coupling, $g_V$. $V^\prime$ is slightly heavier than $V$ with a mass splitting $m_{V^\prime}-m_V\simeq 3.5$~keV.
   The decay of $V^\prime$ to $V$ and a photon gives rise to the 3.5~keV line. The production of $V$ and $V^\prime$ takes place in the early
    universe within the freeze-in framework through the effective $g_V$ coupling when $m_{V^\prime}<T<\Lambda $, $\Lambda$ being the cut-off
     above which the effective $g_V$ coupling is not valid. We introduce a high energy model that gives rise to the $g_V$ coupling at low energies.
      To do this, $V$ and $V^\prime$ are promoted to gauge bosons of   spontaneously broken new $U(1)_V$ and $U(1)_{V^\prime}$ gauge symmetries,
       respectively. The high energy sector includes  milli-charged chiral fermions that lead to the $g_V$ coupling at low energy via  triangle
        diagrams.
\end{abstract}
\section{Introduction}
Although strong hints for the existence of a form of Dark Matter (DM) consisting above 25 \% of the whole energy density of the Universe is established,
 our knowledge on properties of the particles making up the DM is very limited. In particular, we still do not   know the values of DM mass,
  spin and lifetime. We do not also know whether DM consists of a single sort of particle or like ordinary matter comes in varieties of elementary
   and composed particles. There is a rich literature suggesting candidates for DM but most of them focus on the simplest scenario with a single stable
    DM candidate with mass of O($100$ GeV) and with spin equal to 0 or 1/2. Possibility of DM  with spin one ({\it i.e.,} vector DM) has been only recently
     attracted attention \cite{Lebedev,us}.

Recently Ref. \cite{Bulbul} has found a photon line at energy of $(3.55-3.57) \pm 0.03$ keV at more than 3 $\sigma$ C.L. in the data collected by XMM-Newton  observatory from 73 galaxy clusters
 distributed in redshifts between 0.01-0.35. The Chandra data on Perseus also confirms this result \cite{Bulbul}. Ref. \cite{Bulbul} carefully analyzes the possibility of  interpreting this line as an atomic
  transition line but according to \cite{Bulbul} such an interpretation does not seem to be likely within the standard picture (see however, Ref.
   \cite{Profumo}).  Moreover no 3.5 keV signal has been found from Virgo. An independent analysis in Ref. \cite{Boyarsky} finds a similar signal from Andromeda galaxy and Perseus cluster.
 One explanation is decaying DM to a photon.  Assuming that most of dark matter today is composed of one form of decaying particle, in order to explain the intensity of the line, the lifetime has to be \cite{Boyarsky}
 \be \label{tauDM}\tau_{DM} =10^{28}-10^{29} ~{\rm sec} \frac{7 ~{\rm keV}}{m_{DM}} . \ee
 However, one should bear in mind that atomic line emission is not conclusively ruled out \cite{Bulbul,Profumo}. Conclusive results can be achieved
after analysis of Astro-H data \cite{Bulbul,Boyarsky}.
Moreover, according to \cite{Signe} the  Chandra X-ray observations of the Milky Way are consistent with the line at 3.5 keV line only with most conservative assumptions on astrophysical sources
 (see, however, Ref. \cite{new}). Ref \cite{dwg} investigates the presence of the line in the stacked spectra of dwarf spheroidal galaxies and finds no signal. Under standard assumption on the galactic dark matter column density, the null signal excludes the dark matter origin of the 3.5 keV line at 4.6 $\sigma$
C.L. However as pointed out in \cite{dwg}, Ref \cite{Bulbul} does
not include the foreground dark matter halo of the milky way itself.
Inclusion of this foreground will alleviate the tension between the
two results. Considering such debates, it is still premature to
claim a solid observation of the line. Nevertheless, there is
already rich literature trying to present a model explaining the
line by DM decay \cite{decay}, annihilation \cite{annihilation} or
axion-like DM conversion \cite{axionlike}.
  Ref. \cite{Cline:2014eaa} suggests a dark matter model  that explains the line by atomic hyperfine transitions in dark atoms.

In the present paper, we propose a scenario in which dark matter is composed of two vector
 bosons $V$ and $V^\prime$ with $m_{V^\prime}-m_V\simeq 3.5$ keV. $V^\prime$ is metastable and decays into
  $V$ and a photon comprising the 3.5 keV line. The other boson $V$ is protected against decay by a $Z_2$ symmetry.
   The decay of $V^\prime$ proceeds via a generalized Chern-Simons interaction term of form
\be \label{gV} g_V \epsilon^{\alpha \beta \mu \nu}F_{\alpha \beta} V_\mu V_\nu^\prime\ . \ee
This is an effective coupling valid below cut-off $\Lambda$. The same interaction can lead to the observed
amount of DM abundance within the freeze-in framework while $m_{V^\prime}<T<\Lambda$ via $s$-channel  $ f \bar{f} \to \gamma^* \to V V^\prime$.
We introduce a high energy model that  includes (milli-)charged chiral fermions with heavy masses  that lead to the effective coupling in (\ref{gV}) through a triangle diagram.

We introduce the dark matter model in section 2 and a UV-completion for the scenario in section \ref{UV}.   Conclusions are summarized in section 4.

\section{The scenario}
In this section, we show that the $V^\prime \to \gamma V$ decay via the generalized Chern-Simons coupling in Eq. (\ref{gV})
can explain the claimed 3.5 keV line. We then show that the same coupling can lead to $V$ and $V^\prime$ production in the early universe with desired
abundance within the freeze-in framework.

In \cite{us}, we proposed a vector DM model with mass of 130 GeV and a generalized Chern-Simons coupling to explain the 130 GeV line that  had been back then claimed to be observed in the Fermi-LAT data from galaxy center \cite{fermilat}. It is tantalizing to invoke similar scenario to explain the 3.5 keV line via $t$-channel annihilation of a pair of vector DM with mass of 3.5 keV to a photon pair. To account for the intensity \cite{Frandsen}, $g_V/m_{V^\prime}$ should be $\sim (0.18-0.38)$GeV$^{-1}$.
Such values of parameters lead to too large monophoton plus missing energy signal at the LHC via $f \bar{f}\to V V^\prime$ and subsequently $V^\prime \to V \gamma$ \cite{us}. The bound from LHC rules out this scenario \cite{Malik:2012sa}\footnote{Using the eXcited Dark Matter mechanism (XDM) is
 another possibility to explain the X-ray line \cite{XDM}. Within the XDM scenarios,
 dark matter particles  can be up-scattered to the heavier state $V^{\prime}$ via  $\langle \sigma(VV\rightarrow
V^{\prime}V^{\prime}) v\rangle\sim 6\times
10^{-6}({g_V^4}/{m_V^2})(v/2000~{\rm km/sec})$. The 3.5 keV X-ray line can be  subsequently produced
 by $V^{\prime}\rightarrow V\gamma$, provided that  $m_{V^{\prime}}-m_V\simeq 3.5$ keV. Using the results of \cite{XDM} and equating the predicted and observed flux from Perseus, we find
  that $g_V/m_V\sim 1~{\rm GeV}^{-1}$. Such  values of parameters are already ruled out by $VV^\prime$ pair production in
   colliders as well as absence of the monochromatic photon line at $E=m_V$ from dark matter
    annihilation via $\langle \sigma(VV \to \gamma \gamma) v\rangle \sim 0.03 m_V^{-2}\sim 12 \mu b ({\rm GeV}/m_V)^{2}$. }.

Let us now take
\be m_{V^\prime}-m_V\simeq 3.5 ~{\rm keV} \ll m_{V^\prime} \label{split}\ . \ee
Moreover, let us suppose that both $V$ and $V^\prime$ have been produced in the early universe.
The decay of non-relativistic $V^\prime$ will then lead to 3.5 keV line.
In order to explain the intensity, relation Eq. (\ref{tauDM}) should be satisfied.  As shown in \cite{us}, the decay rate is given by:
 \be
 \Gamma(V^{\prime}\longrightarrow V+
 \gamma)=\frac{g_V^2}{24}\frac{\cos^2\theta_W}{\pi}\frac{(m_{V^{\prime}}^2-m_V^2)^3(m_{V^{\prime}}^2+m_V^2)}{m_V^2m_{V^{\prime}}^5}.\nonumber
\ee Assuming that the densities of $V$ and $V^\prime$ are equal today, the lifetime of $V^\prime$ should be half what is shown in Eq. (\ref{tauDM}) to account for the observed intensity of the 3.5 keV line. Thus, we find: \be \label{gVmV} g_V\simeq (5\times10^{-16}-1.5\times10^{-15})
(m_V/{\rm GeV})^{3/2}\ .\ee This coupling is too small to lead to
observable effects at collider experiments. {Through the
generalized Chern-simons coupling, the dark matter can interact
inelastically with nuclei via a $t$-channel photon exchange
\cite{us,Gondolo}. Taking $m_{V^\prime}-m_V\simeq 3.5 ~{\rm keV}$, for
$m_V=30 ~{\rm GeV}$, we find  that the  DM-nucleon cross section should be smaller than $6.2\times 10^{-53}$cm$^2$ which is well below the
bound from LUX \cite{LUX}. 
}

 The production of $V$ and $V^\prime$ will take place via $f\bar f \to \gamma^* \to VV^\prime$ at low temperatures
 $T<\Lambda$ where effective $g_V$ coupling is valid and $\Lambda$ is the cutoff above which the effective $g_V$ coupling is not valid.
 In \cite{us}, the cross section of  $f\bar f \to \gamma^* \to VV^\prime$ is calculated:
 \be
\sigma (f \bar{f} \to V V^\prime)= \frac{(eg_VQ_f \cos\theta_W)^2}
 {12\pi
 N_c E_{cm}^6m_V^2m_{V^\prime}^2}{\mathcal{K}}\mathcal{S}(E_{cm},m_V,m_{V^\prime})\label{ffbar}
 \ee
 where
${\mathcal{K}}=\sqrt{(E_{cm}^2+m_V^2-m_{V^\prime}^2)^2-4m_V^2E_{cm}^2}$
and
$$
\mathcal{S}(E_{cm},m_V,m_{V^\prime})=[E_{cm}^4+(m_V^2-m_{V\prime}^2)^2](m_V^2+m_{V\prime}^2)-2E_{cm}^2(m_V^4-4m_V^2m_{V\prime}^2+m_{V^\prime}^4).
$$
Notice that as $E_{cm} \to \infty$, the production cross section
converges to a constant value. This behavior reflects the fact that
$g_V$ is valid only below $E_{cm}\stackrel{<}{\sim} \Lambda$.
Because of this behavior, most of
  $V V^\prime$ production will take place at high temperatures when  $T\gg m_V$. Using the formulations for freeze-in framework  developed in \cite{Blennow:2013jba} we find
  $$(\Omega_V+\Omega_{V^\prime}) h^2\simeq  1.5\times 10^{22}{g_V^2}\frac{T_f}{m_V}$$
  where  we have set the upper bound  of integration on temperature equal to $T_f$. If the reheating temperature  is smaller than $\Lambda$, we should set $T=T_R$; otherwise, we should set $T_f =\Lambda$. Inserting $g_V^2 \sim 10^{-30} m_V^3/{\rm GeV}^3$  and setting $\Omega_{DM}h^2=(\Omega_V+\Omega_{V^\prime})h^2=0.1$, we obtain \be T_f\simeq 7 \times 10^6~{\rm GeV} \left( \frac{\rm GeV}{m_V}\right)^2 \left( \frac{10^{-15}}{g_V}\right)^2  . \label{Lambda} \ee
Notice that the results depend  on the upper bound  of integration on temperature, $T_f$. This is not unexpected within freeze-in framework. We should however study the  high energy model that  leads to the effective $g_V$ coupling in low temperatures to make sure that at higher temperatures, there is no mechanism to overproduce DM.
This will be done in the next section.
\section{Ultraviolate completion of the scenario \label{UV}}
In the previous sections, we focused on low energies and temperatures at
which the couplings of vector bosons $V$ and $V^\prime$ to SM is
through the generalized Chern-Simons coupling, $g_V$. As we
discussed earlier, the $g_V$ coupling is an effective coupling valid
only below a certain energy scale. In this section, we try to first
introduce a UV-completed model that, below a certain energy scale,
yields the effective coupling in Eq. (\ref{gV}). We then estimate the abundance
of $V$ and $V^\prime$ produced in the early universe at temperatures above the cut-off of the effective coupling. As shown in
\cite{Anastasopoulos:2006cz}, generalized Chern-Simons coupling  can result from triangle
diagram in which chiral fermions propagate. The interesting point is
that $g_V$ turns out to be independent of the masses of these
particles. In the following, we assume that $V$ and $V^\prime$ are
gauge bosons of new $U_V(1)$ and $U_{V^\prime}(1)$ symmetries and
acquire mass via Higgs mechanism. We moreover add Dirac fermions
$\psi_i$ which are colorless and singlet under electroweak SU(2) but
have nonzero hypercharge and therefore nonzero electric charge. The new fermions
are also taken to be in doublet representations of $U_V(1)$ and $U_{V^\prime}(1)$.
That is under $U_V(1)$
$$ \psi_{iR} \stackrel{U_V(1)}{\Longrightarrow} e^{i Q_{iR}\sigma_1} \psi_{iR}    \ \ \ \ {\rm and} \ \ \ \   \psi_{iL} \stackrel{U_V(1)}{\Longrightarrow} e^{i Q_{iL}\sigma_1} \psi_{iL} $$
and  under $U_{V^\prime}(1)$
$$ \psi_{iR} \stackrel{U_{V^\prime}(1)}{\Longrightarrow} e^{i Q^{\prime}_{iR}\sigma_1} \psi_{iR}    \ \ \ \ {\rm and} \ \ \ \   \psi_{iL} \stackrel{U_{V^\prime}(1)}{\Longrightarrow} e^{i Q^{\prime}_{iL}\sigma_1} \psi_{iL},$$
where $\sigma_1$ is  the two by two Pauli matrix. The reason we choose
doublet representation is to maintain the $Z_2$ symmetry that
prevents $V$ and $V^\prime$ from mixing with photon. We will return
to this point later. The Lagrangian of the fermions can be written
as \be \mathcal{L}_{\psi}=\sum_i \left[ \bar{\psi}_{iL} i
\slashed{D}\psi_{iL} + \bar{\psi}_{iR} i \slashed{D}\psi_{iR}+(Y_i
\psi_{iR}^\dagger \Delta_i \psi_{iL}+H.c.) \right], \label{psi}\ee
where  $D_\mu=\partial_\mu-i e_VQ_{i}V_\mu\sigma_1-i e_V^\prime
Q^{\prime}_{i}V_\mu^\prime\sigma_1-ieq_i B_\mu/\cos \theta_W$. The fields $\Delta_i$ are two
by two matrices of scalars which are electrically neutral and
transform under new gauge symmetries as follows: \be \Delta_i
\stackrel{U_V(1)}{\Longrightarrow} e^{iQ_{iR}\sigma_1} \Delta_i
e^{-iQ_{iL}\sigma_1}  \ \ \  \ {\rm and}  \ \ \ \  \Delta_i
\stackrel{U_{V^\prime}(1)}{\Longrightarrow}
e^{iQ^{\prime}_{iR}\sigma_1} \Delta_i e^{-iQ^{\prime}_{iL}\sigma_1}.
\label{Dtransformation} \ee Let us take the potential of $\Delta_i$
as follows\\
\be V(\Delta_i)=-m_{\Delta_i}^2 Tr[\Delta_i^\dagger \Delta_i]
-m_{\Delta_i^\prime}^2 Tr[\Delta_i^\dagger
\sigma_1\Delta_i\sigma_1]+\lambda (Tr[\Delta_i^\dagger
\Delta_i])^2\nonumber\ee \be
+ \lambda_1 \left|Tr[\Delta_i^\dagger
\Delta_i\sigma_1]\right|^2+\lambda_2
\left|Tr[\Delta_i\Delta_i^\dagger \sigma_1]\right|^2\ .  \label{VD} \ee
It is straightforward to show that $V(\Delta_i)$ is invariant under
any \be \Delta_i \to e^{i\alpha_i \sigma_1} \Delta_i e^{i\beta_i
\sigma_1}. \label{Gtransformation}\ee Notice  that transformations
in Eq. (\ref{Dtransformation}) are a subgroup of transformations in
Eq. (\ref{Gtransformation}).  The potential in Eq. (\ref{VD}) is not
the most general potential invariant under (\ref{Gtransformation}).
Combinations such as Tr$[\Delta_i^\dagger \Delta_i \sigma_1]$  also
respect (\ref{Gtransformation}). Our aim here is not to write down
the most general Lagrangian. On the contrary we want to write a
simple Lagrangian that breaks $U_V(1) \times U_{V^\prime}(1)$ in a
desired way maintaining the $Z_2$ symmetry that guarantees the
stability of $V$ and the meta-stability of $V^\prime$ and gives the same mass to the two components of $\psi_{i L}$ doublets. It is
straightforward to verify that
\begin{eqnarray}
\langle \Delta_i \rangle =\left[ \begin{matrix} v_i & 0 \cr 0 & v_i
\end{matrix} \right]\end{eqnarray} is a minimum of the potential in
Eq. (\ref{VD}) and any other minimum can be transformed to this form
by employing transformations (\ref{Gtransformation}). From now on,
we will work in this basis. Notice that the two components of
$\psi_i$ are degenerate with masses \be \label{mpsi} m_{\psi_i}=Y_i
v_i.\ee Moreover the Lagrangian respects a $Z_2$  symmetry under
which $V_\mu$, $V_\mu^\prime$, second component of $\psi_i$,
$(\Delta_i)_{12}$ and $(\Delta_i)_{21}$ are odd but the rest of
fields are even. This $Z_2$ symmetry stabilizes the $V$ boson. Vacuum
expectation values of $\Delta_i$ spontaneously break $U_V(1)$ and
$U_{V^\prime}(1)$ and  induce mass terms for $V_\mu$ and
$V_\mu^\prime$. We will discuss this after fixing the charges.

\begin{table}[htb]
\begin{center}
\begin{tabular}{|c|c|c|c|c|}
\hline ~ & $\psi_{1L}$ & $\psi_{1 R}$ & $\psi_{2L}$ & $\psi_{2 R}$
\cr \hline $U_V(1)$ & 1 &-1 & 1 & 1 \cr $U_{V^\prime}(1)$ & 1 & 1 &
-1 & 1 \cr $U_{em}(1)$ & q & q & q & q \cr \hline
\end{tabular}
\caption{$U(1)$ charges of new fermions.}
\end{center}
\end{table}

Like the case of the standard model $Z$ boson, we can choose a gauge
that  $V$ and $V^\prime$ have three degrees of freedom including
longitudinal components. Notice that masses of $\psi_i$ come from
Yukawa couplings with $\Delta_i$ which are electroweak SU(2) singlets. Since they
develop VEV, they have to be electrically neutral. As a result,
hypercharges of $\psi_{iL}$ and $\psi_{iR}$ should be equal. This automatically
cancels out all anomalies of the $SU(3)\times SU(2) \times U(1)$ gauge group of
standard model. Anomalies of $U_V(1) \times U_{V^\prime}(1)$
symmetries also cancel because we have chosen the doublet
representation and Tr$[\sigma_1 \sigma_1 \sigma_1]=0$. As shown in
\cite{Anastasopoulos:2006cz}, the effective coupling can be written
as
\be \label{gVcombination} g_V=\frac{e e_V e^\prime_V}{48 \pi^2} \sum_i (Q_{iL}
Q^{\prime}_{iR}- Q_{iR} Q^{\prime}_{iL})q_i.\ee Notice that as long
as the two components of the $\psi_i$ doublets are degenerate,
 the amplitude of triangle diagram contributing to $g_V$ in which these two components propagate is equal to what calculated in \cite{Anastasopoulos:2006cz}
 for fermions in singlet representation of $U(1)$. The only difference is that for each doublet the contribution to $g_V$ should be doubled because there are two equal triangle diagrams corresponding to the case that either of two components couple to the photon.
  The  combination of the charges  in Eq. (\ref{gVcombination}) has to be nonzero; however, from anomaly cancelation we find that certain other combinations of the charges must vanish:\\
 (1) Cancelation of the $U_V(1)-U_V(1)-U_{em}(1)$ anomaly implies
 $$ \sum_i (Q_{iL}^2- Q_{iR}^2)q_i=0.$$
 (2) Similarly, cancelation of the $U_{V^\prime}(1)-U_{V^\prime}(1)-U_{em}(1)$ anomaly implies
 $$ \sum_i (Q^{\prime 2}_{iL}- Q^{\prime 2}_{iR})q_i=0;$$
(3) Finally, cancelation of the $U_{V}(1)-U_{V^\prime}(1)-U_{em}(1)$
anomaly implies
 $$ \sum_i (Q^{\prime}_{iL}Q_{iL}- Q_{iR}Q_{iR})q_i=0.$$
 Satisfying all these conditions and obtaining a nonzero $g_V$ is not a trivial problem. In fact, it is straightforward to show that with only a single $\psi_i$ this cannot be done and to obtain a nonzero $g_V$ in a anomaly free theory,  the number of $\psi_i$  has to be increased at least to two. In table 1, we show an assignment of charges for two fermions $\psi_1$  and $\psi_2$ that satisfies all these conditions.

  VEVs of  $\Delta_1$ and $\Delta_2$ will  induce masses for $V$ and $V^\prime$. In general, a tree level mass mixing between $V$ and $V^\prime$ can appear but with charge assignment that we have chosen, no mixing between $V$ and $V^\prime$ appears. The gauge bosons obtain masses as follows
   \be \label{mVVprime} m_V=2 e_V v_1 \ \ \ \ \ {\rm and} \ \ \ \ \  m_{V^\prime}=2 e_{V}^\prime v_2 .\ee  To explain the quasi-degeneracy of  $V$ and $V^\prime$, we can impose
an approximate exchange symmetry on $V(\Delta)$ under $\Delta_1
\leftrightarrow \Delta_2$. This exchange symmetry can be softly
broken by $m_{\Delta}^2$ terms.

Notice that $\psi_1$ and $\psi_2$ having different quantum numbers
cannot mix before breaking $U_V(1)$ and $U_{V^\prime}(1)$.  Vacuum
Expectation Values (VEV) of $\Delta_1$ and $\Delta_2$ however break
$U_V(1)$ and $U_{V^\prime}(1)$, respectively. In principle, mass
terms of $\psi_{1R}^Tc\Delta_2 \psi_{2R} $ and
$\psi_{1L}^Tc\Delta_1\psi_{2L} $ can mix these two. We can however
forbid such terms by imposing a new flavor symmetry under which
$\psi_1 \to \psi_1$ and $\psi_2 \to -\psi_2$. In general, Eq.
(\ref{psi}) induces $V$ and $V^\prime$ mixing at one loop level.
However, as long as we forbid $\psi_1$ and $\psi_2$ mixing, with
particular charge assignment shown is table, no mixing between $V$
and $V^\prime$ appears at loop level. This can be observed by
rewriting gauge couplings of $\psi_1$ and $\psi_2$ in  Eq.
(\ref{psi}) for charge  assignments in table as follows \be (e_V
\bar{\psi}_1 \gamma^\mu \gamma^5 \sigma_1\psi_1 V_\mu+e_{V^\prime}
\bar{\psi}_1 \gamma^\mu  \sigma_1\psi_1 V_\mu^\prime) +( e_V \bar{\psi}_2
\gamma^\mu\sigma_1  \psi_2 V_\mu+e_{V^\prime} \bar{\psi}_2 \gamma^\mu
\gamma^5 \sigma_1\psi_2 V_\mu^\prime). \ee If we forbid the mixing of
$\psi_1$ and $\psi_2$ as well as the mixing of $\Delta_1$ and
$\Delta_2$, the loops will involve either $\psi_1$ or $\psi_2$. As
long as only $\psi_1$ is involved we take $V$ and $V^\prime$
respectively C-odd  and C-even so the $V-V^\prime$ mixing will be forbidden
by charge conjugation symmetry. The operator $\epsilon^{\mu \nu
\alpha \beta} F_{\mu \nu} V_\alpha V_\beta ^\prime$ is however
invariant under charge conjugation and will be therefore allowed.

So far we have not determined the value of the electric charge of
new fermions. In principle, like the millicharge scenario \cite{milli}, $q$ can
be much smaller than 1.  Under global transformations $\psi_1 \to e^{i \alpha_1} \psi_1$ and $\psi_2 \to e^{i \alpha_2} \psi_2$, the Lagrangian in Eq. (\ref{psi}) is invariant. This global $U(1)\times U(1)$ symmetry protects $\psi_1$ and $\psi_2$ from decay. If $\psi_1$ and $\psi_2$ are not produced in the first place, they cannot contribute to the DM abundance. Moreover they cannot lead to production of $V_\mu$ and  $V^\prime_\mu$ via $\bar{\psi}_i \psi_i \to V^{(\prime)}  V^{(\prime)} $.  The following two regimes can be distinguished:
\begin{itemize} \item $T_R<m_\psi$: Obviously, if the reheating temperature is below $m_\psi$, these fermions cannot be produced. The electric charge of $\psi_i$, $q$, can take any value
in the perturbative range. Moreover, we should set $T_f$ in Eq. (\ref{Lambda}) equal to $T_R$. From Eqs. (\ref{mpsi},\ref{gVcombination},\ref{mVVprime}), we therefore find
$$m_V \stackrel {>}{\sim} 2~{\rm GeV} q^{-2/9} (10^{-15}/g_V)^{-8/9} Y^{-4/9}.$$
Taking $g_V=10^{-15} (m_V/{\rm GeV})^{3/2}$, we obtain $m_V>1.5~{\rm GeV} q^{-2/21}Y^{-4/21}$ and $T_R=T_f<10^6~{\rm GeV} q^{10/21} Y^{20/21}$. Taking $m_\psi >T_R>$~TeV, we find $qY^2>5\times 10^{-7}$.
\item $T_R>m_\psi$: In this case, $q$ should be small enough not to lead to production of $\psi_1$ and $\psi_2$ in the early universe.
Taking the production rate in the early
universe to be of order of $\Gamma_\psi \sim e^4 q^4 T/(4 \pi)$, we
find that as long as \be \label{eq} eq <4\times 10^{-5} (m_\psi/{\rm GeV})^{1/4},\ee
the production of $\psi$ in the early universe is negligible\footnote{Notice that such small  electric charge escapes bounds from LEP, LHC and other searches \cite{Dolgov} and can be as light as GeV or even lighter.
}
 ({\it i.e.,} $\Gamma_\psi H|_{T=m_\psi} \ll 1 $). As discussed before, $T_f$ in Eq (\ref{Lambda}) should be set equal to $\Lambda=m_{\psi_i}=v_iY_i$. From Eqs. (\ref{mpsi},\ref{gVcombination},\ref{mVVprime},\ref{eq}), we therefore find
 $$m_V>  7~{\rm GeV} (g_V/10^{-15})^{-3/8} Y^{-1/2}\ . $$ Taking $g_V\simeq 10^{-15} (m_V/{\rm GeV})^{3/2}$, we find $m_V>3.5~{\rm GeV}Y^{-8/25}$ and $m_\psi \sim T_f=13~{\rm TeV}$ and therefore $q<10^{-3}$.
 The reheating temperature can  have any value above  $14$ TeV which is consistent with the canonical picture.
\end{itemize}

\section{Conclusions}
We have presented a dark matter model explaining the 3.5~keV line observed in the XMM-Newton observatory data on galaxy clusters. The model is composed of a light and a heavy sector with a large mass
gap of more than four orders of magnitudes. The light sector includes two vector bosons $V_\mu$ and $V^\prime_\mu$ which play the role of the dark  matter.  The model respects a $Z_2$ symmetry under which standard  model particles are even but $V$ and $V^\prime$ are odd. $V$, being the lightest $Z_2$-odd particle, is stabilized by the $Z_2$ symmetry.
These two vectors couple to the photon via a generalized Chern-Simons coupling, $g_V$. Through this coupling, $V^\prime$ can decay to $V$ and a photon comprising the 3.5 keV line.
The intensity of the line determines the value of the $g_V$ coupling: $g_V\simeq 10^{-15} (m_V/{\rm GeV})^{3/2}$. Such a small coupling cannot lead to an observable signal in the dark matter direct detection experiments or at colliders in any foreseeable future.
The same coupling can however produce enough DM in the early universe via $f \bar{f} \to \gamma^* \to V V^\prime$ within freeze-in mechanism. The production is most efficient at higher temperatures. That is because for energies much larger than $m_{V^\prime}$, the production cross section converges to a constant value given by $g_V^2/m_{V^\prime}^2$. The generalized Chern-Simons coupling is an effective coupling valid only below some cut-off energy, $\Lambda$.  At temperatures above $\Lambda$, the production mechanism
for $V$ and $V^\prime$ should be reconsidered. This means the low energy model including $V$ and $V^\prime$ should be embedded within a UV-completed model that gives rise to the generalized Chern-Simons couplings after integrating out the heavy states. The heavy sector of our model is introduced for this purpose.

The heavy sector includes chiral fermions which are electrically charged. To make the model consistent and renormalizable,
 $V$ and $V^\prime$ are promoted as the gauge bosons of new $U_V(1)$ and $U_{V^\prime}(1)$ symmetries. The new fermions are
  also charged under $U_V(1)$ and $U_{V^\prime}(1)$ and through a triangle diagram  give rise to the $g_V$ coupling
\cite{Anastasopoulos:2006cz}. To maintain the $Z_2$ symmetry that
protects the dark matter against decay, the new fermions are taken
in the doublet representation of the $U_V(1)$ and $U_{V^\prime}(1)$
symmetries. Assigning   charges to these new fermions in a way that
cancels the anomalies of the  $U_Y(1)\times U_V(1) \times
U_{V^\prime}(1)$ symmetry and at the same time yields a nonzero
$g_V$ is a nontrivial exercise and requires at least two generations
of heavy fermions. The Lagrangian of the new fermions in our model
enjoys an accidental remnant global $U(1)\times U(1)$ symmetry that
prevents their decay. Since these particles are electrically
charged, we should make sure that they are not produced in the early
Universe. As we have discussed in detail, this can be realized
within the following two scenarios: (1) The reheating temperature is
below new fermion masses. (2) The electric charges of the new
fermions are too small to let them be produced.

 In {\it summary}, our model contains two relatively light vector bosons that play the role of DM. These vector bosons couple to the photon through a generalized Chern-Simons
 term. This coupling produces the DM particles in the early universe within the freeze-in framework. One of the vector bosons can decay to the other boson and a photon that leads to a detectable monochromatic photon signal from galaxies and galaxy clusters. We have presented a UV-completion of the model leading to the effective Chern-Simons coupling at low energies.

 \section*{Acknowledgements}
The authors thank M. M. Sheikh-Jabbari for useful comments.
 Y.F.  acknowledges partial support from the  European Union FP7  ITN INVISIBLES (Marie Curie Actions, PITN- GA-2011- 289442).

\end{document}